%% file: energy_eff_mu_mimo_asc.tex
\documentclass[conference]{IEEEtran}

\ifCLASSINFOpdf
\usepackage[pdftex]{graphicx}
\else
\fi
\usepackage{epsfig}
\usepackage{epstopdf}
\usepackage{subcaption}
\usepackage{psfrag}

\usepackage[cmex10]{amsmath}
\usepackage{amssymb}

\usepackage{algorithmic}
\usepackage{algorithm}

\usepackage{url}
\usepackage{color}
\usepackage[nolist]{acronym}%[printonlyused]
\usepackage{multirow}

\begin{document}

% Enable bibliography control
\bstctlcite{IEEEexample:BSTcontrol}

\title{Energy and Spectral Efficiency Gains From Multi-User MIMO-based Small Cell Reassignments}

% \author{\IEEEauthorblockN{Michael Shell}
% \IEEEauthorblockA{School of Electrical and\\Computer Engineering\\
% Georgia Institute of Technology\\
% Atlanta, Georgia 30332--0250\\
% Email: http://www.michaelshell.org/contact.html}
% \and
% \IEEEauthorblockN{Homer Simpson}
% \IEEEauthorblockA{Twentieth Century Fox\\
% Email: homer@thesimpsons.com}
% \and
% \IEEEauthorblockN{James Kirk\\ and Montgomery Scott}
% \IEEEauthorblockA{Starfleet Academy\\
% San Francisco, California 96678-2391\\
% Telephone: (800) 555--1212\\



\author{\IEEEauthorblockN{Danny Finn\IEEEauthorrefmark{1},
Hamed Ahmadi\IEEEauthorrefmark{1},
Rouzbeh Razavi\IEEEauthorrefmark{2},
Holger Claussen\IEEEauthorrefmark{2} and
Luiz DaSilva\IEEEauthorrefmark{1}}
\IEEEauthorblockA{\IEEEauthorrefmark{1}CONNECT,
Trinity College Dublin,
Ireland
}
\IEEEauthorblockA{\IEEEauthorrefmark{2}Bell Laboratories, Alcatel-Lucent, Dublin, 
Ireland
}}


\maketitle

\begin{abstract}

In this work we investigate the reassignment of \acp{ue} between adjacent small cells to concurrently enable spatial multiplexing gains through \ac{mu-mimo} and reductions in energy consumption though switching emptied small cells to a sleep state.
We consider a case where \acp{ue} can be reassigned between adjacent small cells provided that the targeted neighbouring cell contains a \ac{ue} with which the reassigned \ac{ue} can perform \ac{mu-mimo} without experiencing excessive multi-user interference, and whilst achieving a minimum expected gain in spectral efficiency over the previous original cell transmissions as a result.
We formulate the selection decision of which \acp{ue} to reassign as a set covering problem with the objective of maximising the number of small cell base stations to switch to a sleep state.
Our results show that, for both indoor and outdoor \acs{lte} small cell scenarios, the proposed \ac{mu-mimo}-based reassignments achieve significant reductions in the required number of active small cell base stations, whilst simultaneously achieving increases in spectral efficiency.

\end{abstract}



% \fi
\IEEEpeerreviewmaketitle

\input{acronyms}

\input{Introduction}
\input{RelatedWork}
\input{SystemModel}

\input{MuMimoAcrossSmallCells}
\input{SimulationScenario}
\input{SimulationResults}
\input{ConclusionsAndFutureWork}

\section*{Acknowledgment}
\vspace{-0.1cm}
We gratefully acknowledge Dr. Andrea F. Cattoni for his much appreciated input in this and our previous work.

This material is based upon works supported by the Science Foundation Ireland under Grants No. 10/CE/I1853 and 10/IN.1/I3007.
\vspace{-0.5cm}
% ************** Removed for submission due to space limitations! Reintroduce for final version! ************** %

 % The authors also acknowledge COST Action IC0902 for funding a short term scientific mission that contributed to this work.
%%%%%%
%%%%%%We also gratefully acknowledge Dr. Rouzbeh Razavi and Dr. Holger Claussen for their input in this work, including the sharing of the Stachus Square pathloss maps.



\bibliographystyle{IEEEtran}
\bibliography{energy_eff_mu_mimo_asc_bib}
% \bibliography{IEEEabrv,../bib/paper}




% \end{thebibliography}

\end{document}

%% file: acronyms.tex
\begin{acronym}
% Instructions on use of acronym list:
%
% When defining a new acronym include it below using the following format:
% \acro{<acronym_tag>}[<ACRONYM>]{<Extended form of acronym>}
%
% When including an acronym in the main text include as
% \ac{<acronym_tag>}
% This will also automatically include the extended form of the acronym on the first occasion it is used.
%
% If you wish to force the acronym to be included in extended form, include as
% \acl{<acronym_tag>}
% For full form:
% \acf{<acronym tag>}
% For short form:
% \acs{<acronym tag>}
% To reset all acronyms
% \acresetall
%
% Full details can be found at:
% ftp://ftp.tex.ac.uk/tex-archive/macros/latex/contrib/acronym/acronym.pdf
%
% KEEP THESE IN ALPHABETIC ORDER!
\acro{3gpp}[3GPP]{3\textsuperscript{rd} Generation Partnership Program}
\acro{awgn}[AWGN]{Additive White Gaussian Noise}
\acro{bicm}[BICM]{Bit-Interleaved Coded modulation}
\acro{bler}[BLER]{BLock Error Rate}
\acro{cdi}[CDI]{Channel Direction Indicator}
\acro{comp}[CoMP]{Coordinated Multi-Point}% transmission and reception}
\acro{csi}[CSI]{Channel State Information}
\acro{cs-rs}[CS-RS]{Cell Specific Reference Signal}
\acro{cqi}[CQI]{Channel Quality indicator}
\acro{cran}[C-RAN]{Cloud Radio Access Network}
\acro{dm-rs}[DM-RS]{Demodulation Reference Signal}
\acro{dr}[DR]{Deployment Ratio}
\acro{earth}[EARTH]{Energy Aware Radio and neTwork tecHnologies}
\acro{eesm}[EESM]{Exponential Effective \ac{sinr} Mapping}
\acro{enb}[eNB]{evolved Node Base station} % / Base station?
\acro{henb}[HeNB]{small cell base station}
\acro{icic}[ICIC]{InterCell Interference Coordination}
\acro{imt-a}[IMT-A]{International Mobile Telecommunications-Advanced}
\acro{irc}[IRC]{Interference Rejection Combining}
\acro{ll}[LL]{Link Level}
\acro{lte}[LTE]{Long Term Evolution}
\acro{lte-a}[LTE-A]{\ac{lte}-Advanced}
\acro{mac}[MAC]{Medium Access Control}
\acro{mcs}[MCS]{Modulation and Coding Scheme}
\acro{miesm}[MIESM]{Mutual Information Effective \ac{sinr} Mapping}
\acro{mimo}[MIMO]{Multiple Input Multiple Output}
\acro{mmse}[MMSE]{Minimum Mean Square Error}
\acro{mno}[MNO]{Mobile Network Operator}
\acro{mu-mimo}[MU-MIMO]{Multi-User MIMO}
\acro{mui}[MUI]{Multi-User Interference}
\acro{nas}[NAS]{Non-Access Stratum}
\acro{nl}[NL]{Network Level}
\acro{pdcp}[PDCP]{Packet Data Convergence Protocol}
\acro{phy}[PHY]{Physical} % Physical Layer
\acro{pmi}[PMI]{Precoding Matrix Indicator}
\acro{rb}[RB]{Resource Block}
\acro{ri}[RI]{Rank Indicator}
\acro{rlc}[RLC]{Radio Link Control}
\acro{roi}[ROI]{Region of Interest}
\acro{rrc}[RRC]{Radio Resource Control}
\acro{rsrp}[RSRP]{Reference Signal Received Power}
\acro{rue}[RUE]{Reassigned \ac{ue}}
\acro{rx}[Rx]{Receive}
\acro{snr}[SNR]{Signal to Noise Ratio}
\acro{sinr}[SINR]{Signal to Interference and Noise Ratio}
\acro{siso}[SISO]{Single Input Single Output}
\acro{sl}[SL]{System Level}
\acro{sic}[SIC]{Successive Interference Cancellation}
\acro{su-mimo}[SU-MIMO]{Single-User MIMO beamforming on a single spatial stream}
\acro{su-mimo2}[SU-MIMO]{Single-User \acs{mimo}}
\acro{sus}[SUS]{Semi-orthogonal User Selection}
\acro{tbs}[TBS]{Transport Block Size}
\acro{ts}[TS]{Technical Specification}
\acro{tti}[TTI]{Transmission Time Interval}
\acro{tue}[TUE]{Target \ac{ue}}
\acro{tx}[Tx]{Transmit}
\acro{ue}[UE]{User Equipment}
\acro{wise}[WiSE]{Wireless System Engineering}
\acro{zf}[ZF]{Zero-Forcing}
\end{acronym}

%% file: Introduction.tex
\section{Introduction}

By significantly improving the spatial reusability of spectrum resources, small cells have shown huge potential to address the capacity deficit that mobile operators are expected to face in the near future. The reduced coverage range of small cell \acp{enb} implies that effective spatial reuse may require dense deployment of such cells, which leads to new radio resource management challenges.

\ac{mu-mimo} is a spatial multiplexing technique in which multiple transmit antennas at the \ac{enb} are used to simultaneously serve multiple \acp{ue} within a single time-frequency resource; this is achieved by transmitting to each \ac{ue} on a different spatial layer. Essentially this consists of directing orthogonal beams at each served \ac{ue}.
Due to \ac{mu-mimo}'s ability to provide large spatial multiplexing gains without requiring additional antennas on the \ac{ue}, and its ability to overcome rank deficiency problems (which often limit single point-to-point spatial multiplexing), \ac{mu-mimo} capabilities have been highly emphasised in recent \ac{3gpp} standardisation.

Often, however, \ac{mu-mimo} is limited to macrocell scenarios. The reason for this is that, in order to keep the interference between simultaneously served \acp{ue} (\ac{mui}) to a minimum, the \acp{ue} must experience sufficiently uncorrelated (semi-orthogonal) channels. In small cell scenarios, where there are few \acp{ue} per cell, it is not always possible to find suitable \ac{ue} sets for \ac{mu-mimo}. However, in dense small cell scenarios a \ac{ue} can often be in range of multiple small cells. By reassigning \acp{ue} between neighbouring \acp{enb} in such a way as to increase the number of suitable \ac{ue} pairings, additional \ac{mu-mimo} spatial multiplexing gains can be enabled \cite{Finn2014}. This can be seen as similar to enlarging the search space for suitable \ac{ue} \ac{mu-mimo} pairings.

\begin{figure}
	\centering
	\begin{subfigure}{0.49\columnwidth}
		\includegraphics[width={\columnwidth}]{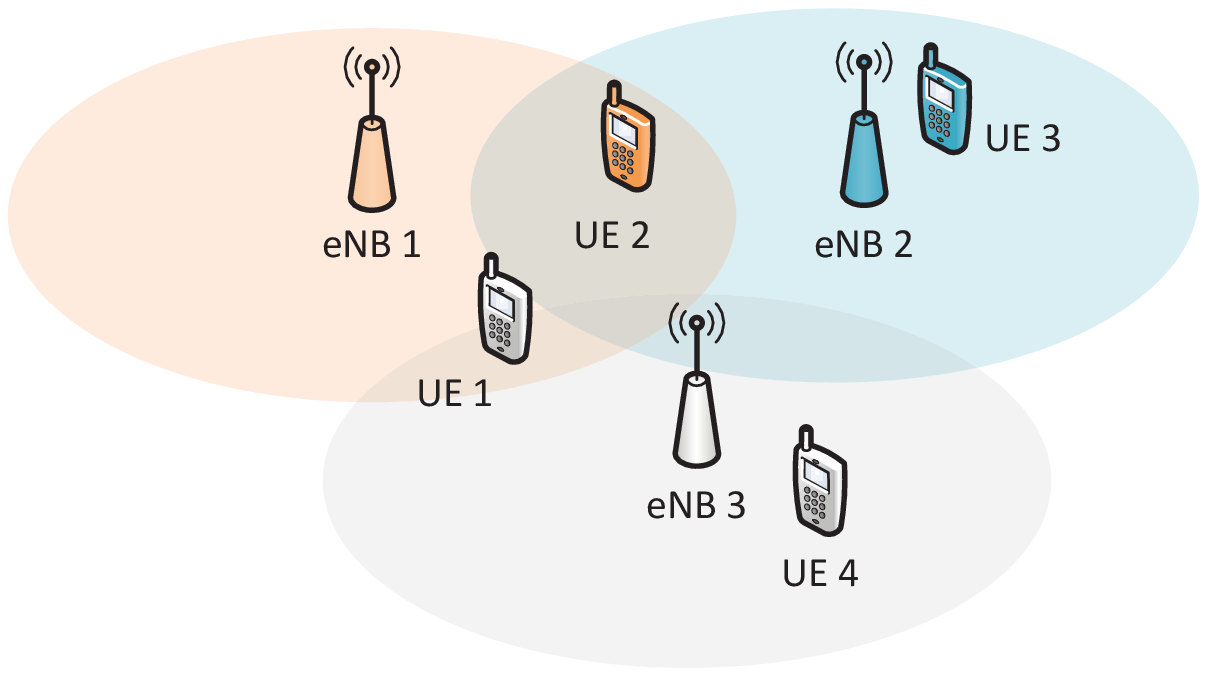}
		\caption{}
		\label{fig:explanation1}
	\end{subfigure}
	\begin{subfigure}{0.49\columnwidth}
		\includegraphics[width={\columnwidth}]{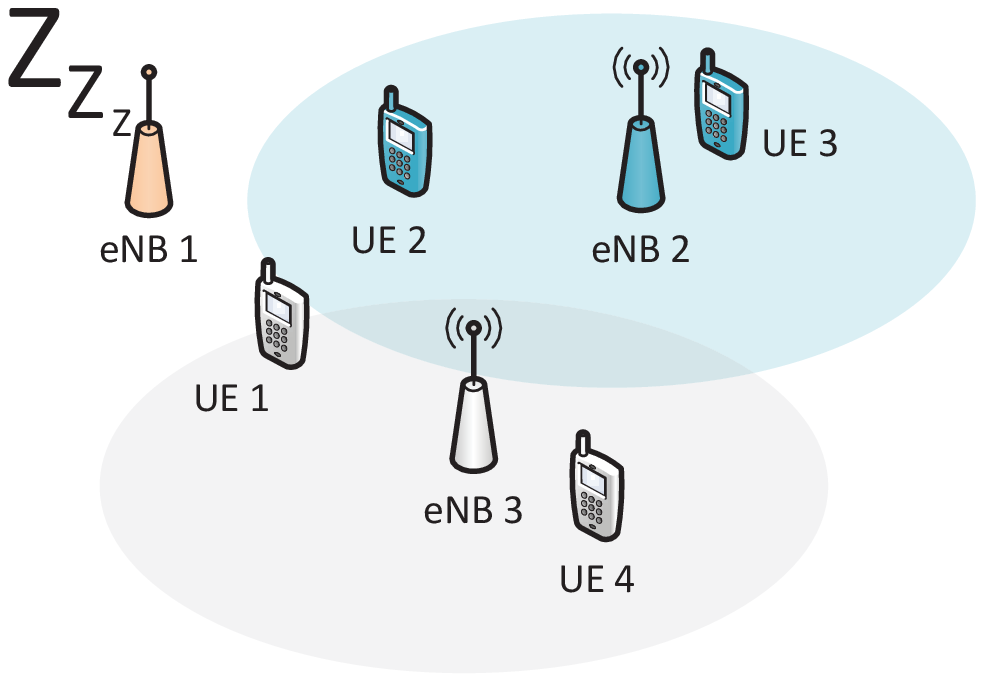}
		\caption{}
		\label{fig:explanation2}
	\end{subfigure}
	\vspace{-0.5cm}
	\caption{\footnotesize UE colours indicate the Small Cell eNBs to which each UE is attached. (a) Consider that by reassigning UE 2 to eNB 2 MU-MIMO transmissions simultaneously serving UEs 2 \& 3 can be enabled (e.g. if UE 2 is within range of eNB 2 and the channels of UEs 2 \& 3 are semi-orthogonal). (b) Additionally, by reassigning UE 2 to eNB 2, as there are no more UEs attached to eNB 1, eNB 1 can be switched to a sleep state to conserve energy.}
	\label{fig:explanation}
	\vspace{-0.8cm}
\end{figure}

Given this scenario, once a \ac{ue} is reassigned to a target neighbouring cell, if the host cell is no longer serving any active \acp{ue}, it can enter an idle state where radio transmissions are temporarily suspended, achieving high energy savings. 
An example of this is illustrated in Figure \ref{fig:explanation}. 
This additionally results in a reduction in interference and pilot pollution problems, which are often observed in dense deployments. Moreover, considering that electricity costs account for 20-30 percent of network operational expenses \cite{Ashraf2011}, there is no lack of motivation for operators to become more energy efficient.

Within this context, this paper introduces a novel scheme where \acp{ue} are selected for reassignment to neighbouring \acp{enb} in order to save energy by reducing the number of active small cell \acp{enb} serving \acp{ue}, while simultaneously achieving spectral efficiency gains by enabling use of additional spatial layers through \ac{mu-mimo}. 
We perform system level simulations to quantify the achievable gains in two scenarios: one indoor and one outdoor.

\vspace{-0.5cm}

%% file: RelatedWork.tex
\section{Related Work}

In our previous work \cite{Finn2014} we proposed a mechanism called \emph{\ac{mu-mimo} Across Small Cells} in which \acp{ue} are reassigned between neighbouring \acp{enb} 
to achieve an increase in spectral efficiency. The current work combines this concept with the use of small cell sleep modes to obtain simultaneous increases in both spectral and energy efficiency by reassigning \acp{ue} in such a way as to concurrently improve \ac{mu-mimo} operation and produce unoccupied small cells, which can then be switched to an idle state to conserve energy.
To reflect this change of emphasis significant changes were required within the reassignment mechanism including 
modification of the method of selection of \acp{ue} to reassign (which is now formulated as a set covering optimisation), 
introduction of a new parameter $\tau$ to grant control over the tradeoff between energy and spectral efficiency,
and considerations concerning sleep state operation.

As discussed in \cite{Ashraf2011}, sleep mode techniques often fall into one of three categories, differentiated by the method used for their reactivation. These are: Small Cell Controlled, in which an RF sniffer is utilised to identify potential \acp{ue} to serve \cite{Ashraf2010}; Core Network Controlled, in which, as in our work, small cells are put into sleep states and re-awoken by a centralised core network element; and finally \ac{ue} controlled, in which \acp{ue} emit periodic wake-up signals for surrounding small cells.

Traditionally these works have taken advantage of low traffic conditions and the energy saving possibilities they present, although more recently, given the density of small cell deployments and the associated cell redundancies/over provisioning, the work has evolved to also include user association considerations \cite{Li2013,Park2013}. 
In \cite{Li2013} centralised decisions are made to switch a portion of the active small cell \acp{enb} to sleep modes, from which the resulting gains in energy efficiency and \ac{sinr} (as a result of reductions in inter-cell interference) are studied through system level simulation. Our work further improves the \ac{ue} performance by selectively deactivating small cell \acp{enb} in such a way that achieves both \ac{mu-mimo} spatial multiplexing and interference reduction spectral efficiency gains.
In \cite{Park2013} uplink \ac{cran} is considered with centralised joint decoding of the received signals of multiple multi-antenna small cell \acp{enb}. 
The authors introduce a sparsity-inducing term (\ac{enb} activity cost) to their optimisation to jointly maximise rate and minimise the number of active small cells.
Unlike \cite{Park2013} we do not consider joint encoding or decoding. 
Instead \ac{mu-mimo} transmissions are performed separately by each small cell \ac{enb}, which considerably reduces computation and coordination overheads.

The concept of coordinating the reassignment of \acp{ue} in order to achieve \ac{mu-mimo} gains can be related to a number of \ac{comp} concepts, most notably Coordinated Beamforming/Scheduling, Network MIMO and Dynamic Cell/Point selection. Our work differs considerably from these by actively reassigning \acp{ue}, avoiding joint transmissions and applying \ac{mu-mimo} considerations, respectively, as well as achieving additional gains through use of sleep states. \cite{Pateromichelakis2013} provides a recent survey on these and other similar multi-cell scheduling concepts.

%% file: SystemModel.tex
\section{System Model}

Our network contains $N_{UE}$ \acp{ue}, each with $N_r$ receive antennas, and $N_{eNB}$ \acp{enb}, each with $N_t$ transmit antennas. The \acp{enb} constitute a heterogeneous mix of tri-sector macrocell \acp{enb} and open subscriber group small cell \acp{enb} all operating within the same frequency band. Our \acp{ue} of interest are those served by the small cell \acp{enb} and as such the macrocell \acp{enb} are simply modelled as sources of inter-cell interference.

The small cell \acp{enb} are governed by a central coordinator which is responsible for initiating reassignments and small cell sleep states.
\acp{ue} are only served by a single \ac{enb} at any point in time. This means coherent transmissions from multiple cells are not necessary and the coordination requirements on the central controller are low relative to other \ac{comp} techniques.

A small cell \ac{enb} $e$ can operate in either an \textit{active} or \textit{idle} state; indicated by $a_e\!=\!1$ or $a_e\!=\!0$, respectively.

\vspace{-0.2cm}

\subsection{Signal Model}
For \ac{mu-mimo} with two \acp{ue} co-scheduled, each transmitting on a single spatial layer (rank-1), the received signal of \ac{ue} $k$, co-scheduled for \ac{mu-mimo} with \ac{ue} $j$, is given by

\vspace{-0.7cm}

\begin{equation}
\label{eqn:channel_model}
\mathbf{y}_{k} = \mathbf{H}_{k,O}\mathbf{w}_{k}x_{k} + \mathbf{H}_{k,O}\mathbf{w}_{j}x_{j} + \!\!\!\! \sum_{l=1}^{N_{eNB}-1} \!\!\!\! a_l \mathbf{H}_{k,l}\mathbf{W}_{l}\mathbf{x}_{l} + \mathbf{n}_{\textup{W},k}
\end{equation}

\vspace{-0.3cm}

\noindent where $\mathbf{y}_{k}$ is the $N_r \!\! \times \! 1$ received signal vector, $\mathbf{H}_{k,O}$ is the $N_r \!\! \times \!  N_t$ channel matrix within the original serving cell, $\mathbf{w}_{k}$ is the $N_t \!\! \times \!  1$ applied unitary precoding and $x_{k}$ is the transmitted symbol, of \ac{ue} $k$. $\mathbf{H}_{k,O}\mathbf{w}_{j}x_{j}$ is the interference from the co-scheduled \ac{ue} $j$, while $\mathbf{H}_{k,l}\mathbf{W}_{l}\mathbf{x}_{l}$ is the interference from neighbouring cell $l$ and finally $\mathbf{n}_{\textup{W},k}$ is complex \ac{awgn}, the elements of which have zero mean and variance $\sigma^2$. It should be noted that the value $\mathbf{H}_{k,O}$ in this equation includes the transmit power to each of the two co-scheduled \acp{ue}, which is one half of the power they would have if they were scheduled alone for \ac{su-mimo}. As indicated by $a_l$, if a neighbouring \ac{enb} is not active it does not produce inter-cell interference. It is assumed that all serving cells are active, hence we omit the $a_O$ which would otherwise precede $\mathbf{H}_{k,O}$.

The post-reception \ac{sinr} of \ac{ue} $k$,
after the application of a $1 \! \times \! N_r$ receive filter $\mathbf{g}_k$, 
can be represented as

\vspace{-0.5cm}

\begin{equation}
\label{eqn:sinr_mu-mimo}
\gamma_{k} = \frac{| \mathbf{g}_{k} \mathbf{H}_{k,O}\mathbf{w}_{k} |^2}
					{| \mathbf{g}_{k}\mathbf{H}_{k,O}\mathbf{w}_{j} |^2 + 
					 | \mathbf{g}_{k} \sum_{l=1}^{N_{eNB}\!-\!1} \!\! a_l \mathbf{H}_{k,l}\mathbf{W}_{l} |^2 + 
					 \sigma^2\mathbf{I} || \mathbf{g}_{k} ||^2}.
\end{equation}

\vspace{-0.3cm}

\noindent We apply a \ac{mmse} with \ac{irc} filter \cite{DelCarpioVega2012}. This filter combines the suppression of co-layer intra-cell interference by the \ac{mmse} filter and suppression of inter-cell interference through \ac{irc} filtering. This receiver is also known as the Advanced \ac{lte} \ac{ue} Receiver and is the baseline receiver from \ac{3gpp} \ac{lte} Rel.11 onward \cite{Holma2012}.

\subsection{CSI calculations} \label{sec:csi_fb}

The \ac{lte} standards specify three types of \ac{csi} feedback for use in the scheduling of \acp{ue} and adaptive modulation and coding. 
The \ac{cqi} is a quantised form of \ac{sinr}, the \ac{pmi} recommends a linear precoding matrix from a predefined codebook, and the \ac{ri} indicates the rank (number of parallel transmission streams) to use \cite{3GPPTS36.213}.
For both \ac{su-mimo} and \ac{mu-mimo} operation, we consider only Rank-1 transmissions for each \ac{ue}.

In this work, we consider wideband \ac{pmi} feedback. The \ac{pmi} is selected as the one which maximises the \ac{su-mimo} mutual information summed over all subcarriers in the channel bandwidth. We consider \ac{lte} Rel.10 \ac{mu-mimo} operation which supports the subband use of \ac{mu-mimo} on some subcarriers and \ac{su-mimo} on others, as well as supporting the use of Zero-forcing to orthogonalise non-orthogonal precoders in \ac{sus}. For 2 or 4 \ac{tx} antenna configurations Rel.10 \ac{mu-mimo} makes use of the Rel.8 \ac{mu-mimo} precoding codebook.

\subsubsection{MU-MIMO CQI calculation} \label{sec:mu-mimo_cqi}

Computing \ac{cqi} feedback for \ac{mu-mimo} scenarios is always a problem as the \acf{mui} is a function of which \acp{ue} are selected to be simultaneously served. Furthermore, if the channel conditions do not suit \ac{mu-mimo} the scheduler may fallback to \ac{su-mimo} operation.
This makes it not possible to characterise the \ac{sinr} prior to the scheduling process.
For this reason, we model the \ac{mu-mimo} \ac{cqi} as proposed in \cite{Nguyen2012}.
All \acp{ue} feed back \acp{cqi} for \ac{su-mimo} (not including \ac{mui}) which are adjusted to account for the splitting of the transmit power between the co-scheduled \acp{ue}, and for the mean level of unsuppressed \ac{mui}, $\Delta_{MUI}$, as follows \cite{Nguyen2012}:

\vspace{-0.6cm}

\begin{equation}
\label{eqn:mu-mimo_cqi}
CQI_{\textup{MU-MIMO}} = \frac{1}{\frac{n_{MU}}{CQI_{\textup{SU-MIMO}}} + \Delta_{MUI}(n_{MU}-1)}
\end{equation}

\vspace{-0.1cm}

\noindent where $n_{MU}\!$ is the number of \acp{ue} co-scheduled for \ac{mu-mimo} within a single time-frequency \ac{rb} and $CQI_{\textup{SU-MIMO}}$ is the fed back quantised \ac{sinr} in linear form.

$\Delta_{MUI}$ is a precomputed parameter, specific to the environment and device configuration. It is obtained by taking the expectation over a large number of channel realisations of the ratio of the suppressed \ac{mui} power to the \ac{su-mimo} signal power (without power splitting between co-scheduled \acp{ue}):

\vspace{-0.2cm}

\begin{equation}
	\label{eqn:delta_mui}
	\Delta_{MUI} = \mathbb{E} \left[ \frac{| \mathbf{g}_{k} \mathbf{H}_{k,O}\mathbf{w}_{j} |^2}{n_{MU}| \mathbf{g}_{k} \mathbf{H}_{k,O}\mathbf{w}_{k} |^2} \right].
\end{equation}

\vspace{-0.2cm}

\noindent A fixed value of $\Delta_{MUI}$ is made reasonable by the constraint that the maximum allowed channel correlation of simultaneously served \acp{ue}, $\epsilon$, is very low. This means that between different effective channel matrix realisation pairs the variance in the amount of suppression that can be achieved is also low.

\subsubsection{Neighbouring eNB Feedback} \label{sec:neighbouring_enb_feedback}

In order to assess the feasibility of potential reassignments, when the reassignment mechanism is initiated the \ac{ue} needs to check the quality of the channels between it and any target neighbouring \acp{enb}. To this end, as detailed later in Section \ref{sec:mu-mimo_across_small_cells}, unpaired \acp{ue} compute a wideband \ac{pmi} and a wideband \ac{cqi} (\ac{cqi} averaged across subbands over the transmission bandwidth using \ac{miesm}, or similar) for the target cell.

In \ac{lte} systems the \ac{cs-rs} offers the primary means of neighbouring cell \ac{csi} estimation. \ac{cs-rs} locations of neighbouring cells are shifted in the frequency-domain based on a $\mod(6)$ of their cell ID \cite{3GPPTS36.211}. This was originally intended to provide frequency-domain orthogonality between the reference signals of neighbouring cells in a macrocell hexagonal grid structure, although the \acp{cs-rs} are also coded with cell-specific Zadoff-Chu sequences to provide orthogonality in denser deployment cases (e.g. small cell scenarios). 
\ac{cqi} values obtained in this way are then adapted for different values of $n_{MU}$ according to Eqn. \eqref{eqn:mu-mimo_cqi}.

In this work, the target cell \ac{cqi} is used only in the reassignment decision, and reassignments are only performed which enable the deactivation of the original cell. Hence, we do not include interference from the original \ac{enb} in the target cell \ac{cqi}. This provides a more accurate estimate of the post-reassignment rate. Nonetheless, it is not possible to know, prior to the reassignment decision, which other neighbouring \acp{enb} will also be deactivated. If additional neighbouring \acp{enb} are deactivated the levels of interference will be further reduced, thus the target cell \ac{cqi} provides a conservative estimate of the post-reassignment rate.

\subsection{\ac{mu-mimo} Spectral Efficiency} \label{sec:spec_eff}

In both the scheduling and reassignment processes it is necessary to compare the expected relative performances of \ac{su-mimo} and \ac{mu-mimo}. For this, we map the $CQI_{\textup{SU-MIMO}}$ and $CQI_{\textup{MU-MIMO}}$ values to expected rates per \ac{rb} based on a fitting to \ac{bicm} curves, as discussed in \cite{Schwarz2010se}. We represent the expected rate of a \ac{ue} $k$, when co-scheduled as one of $n_{MU}$ \acp{ue}, attached to \ac{enb} $e$ as $r_{n_{MU},k,e}$. In this work $e$ will be either the currently attached \ac{enb}, which we will call the original \ac{enb}, $O$, or a neighbouring \ac{enb} targeted for reassignment, $T$.

When \ac{mu-mimo} is used ($n_{MU}\!\!\!\!\geq\!\!\!2$), a single \ac{rb} is shared amongst multiple co-scheduled \acp{ue}, each on different spatial layers; as the \acp{ue} each only part-occupy the \ac{rb} bandwidth, their \ac{mu-mimo} spectral efficiency in bits per \ac{rb} is given by their rate for the \ac{rb}, multiplied by the number of simultaneously served \ac{ue} that the \ac{rb} is divided between.

\subsection{Energy Consumption Model} \label{sec:energy_eff}

We model \ac{enb}  input power consumption according to the framework set out by the FP7 \acs{earth} Project \cite{Imran2010} as follows:

\vspace{-0.1cm}

\begin{equation}
	\label{eqn:power_consumption}
	P_{in} = \left\{ \begin{array}{c c}
				N_{t} (P_0 + \Delta_P P_{out}),	&	ACTIVE	\\
				N_{t} P_{sleep},				&	IDLE	\\
			 \end{array} \right.
\end{equation}

\vspace{-0.1cm}

\noindent where $P_0$ represents the power consumption at zero RF output power, $P_{out}$ represents the load-dependent RF output power which has a maximum of $P_{max}$ and slope $\Delta_P$. For full buffer traffic, as in this work, transmissions will operate at maximum power, meaning that $P_{out}$ for active \acp{enb} will be $P_{max}$. $P_{sleep}$ represents the power consumed by an \ac{enb} in sleep mode. An \ac{enb} can switch into a sleep mode to save energy if there are no \acp{ue} requesting transmissions which it has to serve. These power consumption values will differ depending on the type of base station considered, be it femto, pico or macro.

%% file: MuMimoAcrossSmallCells.tex
\section{Reassignment Mechanism} \label{sec:mu-mimo_across_small_cells}

The \emph{\ac{mu-mimo} across Small Cells} reassignment mechanism of this work follows a common structure to our previous work \cite{Finn2014}, augmented by the deactivation of idle small cells, and consists of five main steps:

\begin{enumerate}
	\item[A.] Selection of Considered \acp{ue}
	\item[B.] Check for Target \acp{ue}
	\item[C.] Selection of \acp{ue} to reassign from Reassignable \acp{ue}
	\item[D.] Reassignment
	\item[E.] Deactivation of emptied small cells.
\end{enumerate}

\noindent Additionally, as necessitated by the energy consumption minimisation objective of the reassignment, the third step, in which \acp{ue} are selected for reassignment, is also added.

\begin{figure}
	\centering
	\includegraphics[width=0.8\columnwidth]{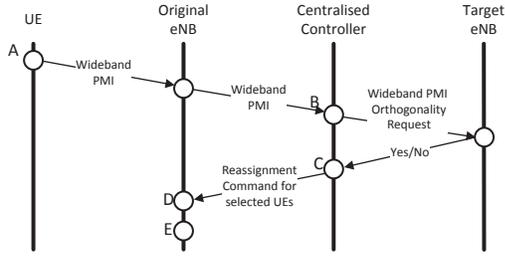} % trim = [l b r t]
	\caption{\footnotesize \ac{mu-mimo} Across Small Cells with sleep mode usage. A. \acp{ue} assess whether they should be considered for reassignment. B. The Centralised Controller checks the target \acp{enb} for potential Target \acp{ue} with which the Considered \acp{ue} can be co-scheduled for \ac{mu-mimo}. If such Target \acp{ue} exist, the corresponding considered \acp{ue} are termed reassignable. C. The Centralised Controller selects a subset of the Reassignable \acp{ue} to reassign. In this paper, these are selected so as to minimise the required number of active \acp{enb}. D. The selected \acp{ue} get reassigned. E. \acp{enb} with no remaining attached \acp{ue} switch to a sleep state to conserve energy.}
	\label{fig:timeline_diagram}
	\vspace{-0.8cm}
\end{figure}

\subsection{Selection of Considered \acp{ue}}

To avoid unnecessary exchange of feedback information, at the start of the reassignment process, \acp{ue} assess whether they should be considered for \ac{mu-mimo}-based cell reassignment.

A \ac{ue} capable of \ac{mu-mimo} operation in its current cell is unlikely to improve its \ac{mu-mimo} capabilities by reassigning to a neighbouring cell 
and is thus
removed from the set of Considered \acp{ue}. 
These removed \acp{ue} meet the constraint

\vspace{-0.4cm}

\begin{equation} \label{eqn:candidate_ues1}
	2\overline{r_{2,k,O}}I_{k,O} > \overline{r_{1,k,O}}.
\end{equation}

\vspace{-0.4cm}

\noindent where $\overline{r_{n,k,e}}$ denotes the expected instantaneous rate averaged over all \acp{rb}. $n\!=\!1$ and $n\!=\!2$ denote SU- and two-layer MU- MIMO, respectively, and the averaged rates are scaled by the number of co-scheduled \acp{ue} (layers) to correspond to the average spectral efficiencies in bits per \ac{rb}. $I_{k,e}$ indicates the presence of another \ac{ue} attached to \ac{enb} $e$ with a precoding matrix semi-orthogonal to that of \ac{ue} $k$.

Of these \acp{ue},
those which expect a predefined minimum gain from \ac{mu-mimo} usage in a neighbouring \ac{enb} are considered for reassignment:

\vspace{-0.4cm}

\begin{equation} \label{eqn:candidate_ues2}	
	2\overline{r_{2,k,T}} - \overline{r_{1,k,O}} > \tau.
\end{equation}

\vspace{-0.4cm}

\noindent where $\tau$ is an offset indicating how much higher the target cell expected rate must be, which can be either positive or negative, and $\overline{r_{2,k,T}}$ is obtained from neighbouring \ac{enb} reference signals as outlined in \ref{sec:neighbouring_enb_feedback}.

The parameter $\tau$ is introduced to provide control over the tradeoff between spectral and energy efficiency. A positive value of $\tau$ indicates a strict requirement on the minimum expected spectral efficiency gains, while a negative value of $\tau$ indicates the willingness to sacrifice some spectral efficiency for a reduction in energy consumption.

\acp{ue} meeting constraints \eqref{eqn:candidate_ues1} and \eqref{eqn:candidate_ues2}
are then labelled the \textit{Considered \acp{ue}}. These Considered \acp{ue} share their wideband \ac{pmi}, for each target neighbouring \ac{enb}, with the central coordinator via their currently attached \ac{enb}.

\subsection{Check for Target UEs}

We term \textit{Target \acp{ue}} as \acp{ue} attached to the target neighbouring \ac{enb} with which the Considered \ac{ue} could potentially be co-scheduled for \ac{mu-mimo}. The central controller checks for Target \acp{ue} by checking if the precoder of any \ac{ue} in the neighbouring cell is semi-orthogonal to that of the considered \ac{ue} (corresponding to the fed back wideband \ac{pmi}), and that their \ac{sinr} is not too low for \ac{mu-mimo} operation to be beneficial. Any Considered \acp{ue} for which a Target \ac{ue} exists are termed \textit{Reassignable \acp{ue}}.

\vspace{-0.2cm}

\subsection{Selection of \ac{ue} to reassign from the Reassignable \acp{ue}}

The choice of \acp{ue} to reassign can determine the number of \acp{enb} which can be deactivated as a result of the reassignment, as well as the increases in spectral efficiency and \ac{mu-mimo} usage that can be achieved.

In this work we formulate this decision as a set covering problem in which we select \acp{ue} to reassign so as to \emph{maximise the number of deactivated \acp{enb}}. In this, the set of all \acp{ue} $k\in\mathcal{U}$ must be covered (served) by the set of active base stations. However a \ac{ue} $k$ can only be served by an \ac{enb} $e\in\mathcal{E}$ if \emph{either} $e$ is $k$'s original \ac{enb} \emph{or} $k$ can be reassigned to $e$.

This set covering problem can be expressed as follows:

\vspace{-0.3cm}

\begin{equation*}
\begin{aligned}
& \text{minimize} 	& \sum_{e\in\mathcal{E}}(a_{e})	& & \\
& \text{subject to} & \sum_{e \in f(k)}(a_{e}) \geq 1 & & \forall k\in \mathcal{U}	\\
& 					& a_{e} \in \{0,1\} & & \forall e \in \mathcal{E}
\end{aligned}
\end{equation*}

\vspace{-0.3cm}

\noindent where $a_{e}$ indicates that \ac{enb} $e$ is active and $f(k)$ indicates the set of \acp{enb} which can serve \ac{ue} $k$.

In order to assess the performance of the mechanism under optimal reassignment selection (maximum number of deactivated \acp{enb}) we solve this problem using the IBM ILOG CPLEX Optimiser.
The set covering problem is NP-hard; however, as our problem size is relatively small, CPLEX finds solutions in a reasonable amount of time. Nonetheless for real implementations a heuristic solution would be required, for which polynomial time heuristics achieving $\Theta(\ln N_{UE})$ approximations exist \cite{Chvatal1979}.

\vspace{-0.2cm}

\subsection{Reassignment}

All \acp{ue} attached to the small cell \acp{enb} selected to switch to sleep mode are reassigned to their respective target \acp{enb}.

\subsection{Deactivation of emptied small cells}

Once these \acp{ue} have been reassigned, the emptied small cell \ac{enb} can enter the idle mode regime to minimise energy consumption and to avoid creating unwanted interference. However, there is a need for effective mechanisms that can detect the presence of a new \ac{ue} and subsequently wake up the base station. This can be done in either a distributed or centralised fashion.
	 
The simplest distributed wake up mechanism is to equip the \ac{enb} with an RF sniffer which monitors detected energy in the uplink band. Once this quantity exceeds a certain threshold, the \ac{enb} can interpret this as a sign indicating the presence of a nearby \ac{ue} \cite{Ashraf2010}. However, setting the detection threshold optimally is difficult and, contrary to the scenario in \cite{Ashraf2010} where the \ac{ue} communicates with an underlying macrocell \ac{enb}, in dense deployment scenarios, the \ac{ue} is more likely to be served by a neighbouring small cell \ac{enb}. This implies that the \ac{ue} transmits at lower power, making detection more difficult.

In the centralised approach a separate network element triggers the small cell \ac{enb} to wake up. This has the advantage that the base station will not be triggered unnecessarily if the \ac{ue} can be served by other existing \acp{enb}. The challenging task is to determine when a \ac{ue} falls under the coverage area of a given small cell base station. This can be done using RF-fingerprints to estimate the \ac{ue}'s location or through other advanced geolocation methods.

As our network already contains a centralised controller the centralised approach presents the best fit to our scenario.
Furthermore, this approach provides a higher energy efficiency gain than the distributed approach, as less infrastructure is required to detect when a reactivation is required \cite{Ashraf2011}.

%% file: SimulationScenario.tex
\vspace{-0.3cm}	

\section{Investigated Scenarios and Settings}

\vspace{-0.1cm}

We explore the potential benefits of the proposed coordinated \ac{mu-mimo}-based reassignment and sleep mode activation mechanism in two small cell scenarios, both subject to external macrocell interference. The first is an indoor residential femtocell deployment, while the second is an outdoor picocell deployment in the commercial centre of a major European city.

\vspace{-0.3cm}

\subsection{Indoor Residential Scenario}

\vspace{-0.1cm}

\begin{figure}
	\centering
	\includegraphics[width={\columnwidth}]{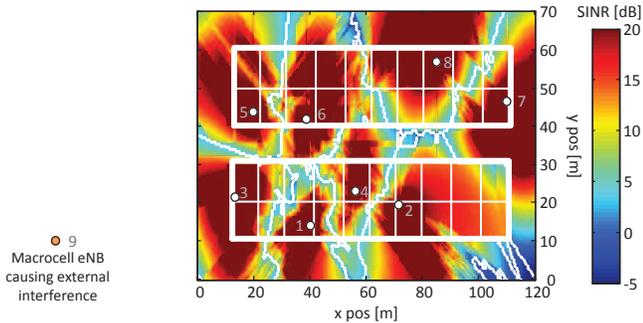}
	\caption{\footnotesize Dual Stripe scenario \ac{sinr} distribution from simulation with a DR of 0.2. White dots mark the HeNB locations, the orange dot marks the location of the external macrocell.}
	\label{fig:simulated_scenario}
	\vspace{-0.8cm}	
\end{figure}

For indoor residential femtocell (HeNB) deployments the \ac{3gpp} Dual Stripe model \cite{3GPPRAN42009} is the most widely used. This model 
consists of two apartment buildings side-by-side, each subdivided into apartments separated by walls. The density of HeNBs deployed is indicated by the \ac{dr}, which denotes the probability of a given apartment containing a HeNB. 
We consider a single-story deployment with a \ac{dr} of 0.2 (8 HeNBs deployed on average). The \ac{sinr} distribution of an example deployment is illustrated in Figure \ref{fig:simulated_scenario}.

\vspace{-0.3cm}

\subsection{Outdoor City Centre Scenario}

\vspace{-0.1cm}

\begin{figure}
	\centering
	\includegraphics[width={0.5\columnwidth}]{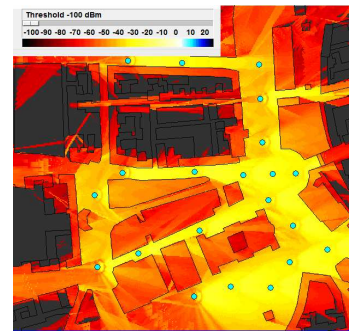}
	\caption{\footnotesize Maximum RSRP map for Munich picocell deployment. The locations of the 21 picocell \acp{enb} are marked by cyan dots. \acp{ue} also experience macrocell interference from 6 surrounding tri-sector base stations the locations of which correspond to those of one of Germany's top-tier mobile operators. }
	\label{fig:munich_scenario}
	\vspace{-0.8cm}	
\end{figure}

We also study the performance of our proposed mechanism in an outdoor scenario, the area surrounding Stachus Square, a commercial zone in the city centre of Munich, Germany.
Figure \ref{fig:munich_scenario} shows the maximum picocell \ac{rsrp} within the 250$m$ x 250$m$ investigated region. 
The detailed 3D ray-tracing tool \ac{wise} \cite{Fortune1995} was used to accurately compute the pathloss and shadow fading in this scenario taking fully into account the locations of the \acp{enb}, building and other obstacles.

\vspace{-0.3cm}

\subsection{Common Parameters/ General Settings}

\vspace{-0.1cm}

All \acp{enb} each have four closely spaced transmit antennas, while the \acp{ue} have two receive antennas which use \ac{mmse}-\ac{irc}. 
At the start of simulation all small cell \acp{enb} operate in an active state and have the same number of allocated \acp{ue}. \acp{ue} are initially assigned to the \ac{enb} with the highest \ac{rsrp}. 
\acp{ue} can only be reassigned between small cells and cannot be reassigned to a macrocell.

Both before and after the reassignments, \acp{ue} are scheduled using \acf{sus} \cite{Yoo2007} modified for proportional fairness. 
This algorithm first selects the \ac{ue} with the highest proportional fair metric to be scheduled, where the proportional fair metric is the ratio of the instantaneous achievable rate to the long-term average throughput of a \ac{ue}. Next a set is computed of the \acp{ue} with quantised channel vectors semi-orthogonal to that of the the first selected \ac{ue}.\footnote{The quantised channel vectors of a \ac{ue} is obtained from the fed back \ac{pmi} as the pseudo-inverse of the recommended precoding vector, and two quantised channel vectors are called semi-orthgonal if the spatial correlation between them is below a predefined bound, $\epsilon$.} 
From this semi-orthogonal set the \ac{ue} with the highest proportional fair metric is selected to be paired with the already scheduled \ac{ue}. Finally, a check is performed to ensure that the proportional fair metric of the two selected \acp{ue} both performing \ac{mu-mimo} exceeds that of the first \ac{ue} alone performing \ac{su-mimo}. Whichever of the two possibilities has the better performance is then used.
To avoid excessive user specific reference signalling overhead and as there are few \acp{ue} per cell, the maximum number of \acp{ue} that can be spatially co-scheduled for \ac{mu-mimo} in a single \ac{rb} is two; however, to support $n\!>\!2$, the mechanism would not require much change.

The parameters used in the energy efficiency model are those provided in \cite{Imran2010} adjusted for the 4 transmit antenna case such that the maximum radiated power is kept consistent.
Additional simulation parameters are provided in Table \ref{tab:sim_params}.

\begin{table}
	\centering
	\caption{Simulation parameters}
	
	\label{tab:sim_params}
	\resizebox{\columnwidth}{!}{
	\begin{tabular}{| l | l | l |}
		\hline \hline
		Scenario Specific Parameters		&	3GPP Dual Stripe (Indoor) 				&	Munich (Outdoor)				\\
		\hline \hline
		Pathloss Model 				&	3GPP Dual Stripe  \cite{3GPPRAN42009}	&	WiSE \cite{Fortune1995}			\\
		\hline
		Fast Fading Model 				&	Winner II \cite{Kyosti2007}				&	Winner II \cite{Kyosti2007}	\\
		\hline
		Deployment Densities			&   \ac{dr} = 0.2							&	Mean inter-site distance = 37 m	\\
		\hline
		Number of Small cells			& 	8 on average							&	21				\\
		\hline
		Small cell maximum Tx Power		& 	20 dBm									&	24 dBm		\\
		\hline
		Number of Macrocell interferers		&	3 										&	18			\\
		\hline
		Macrocell maximum Tx Power		&	46 dBm									&	43 dBm		\\
		\hline
		$P_{max}$					&	0.025 W									&	0.065 W		\\
		\hline
		$P_0$						&	2.4 W									&	3.4 W				\\
		\hline
		$\Delta_P$					&	4.0										&	8.0			\\
		\hline
		$P_{sleep}$					&	1.45 W									&	2.15 W		\\
		\hline \hline
								& 	\multicolumn{2}{|c|}{General Parameters}								\\
		\hline \hline	
		Bandwidth					&	\multicolumn{2}{|c|}{10 MHz}										\\
		\hline
		Snapshot length	 			& 	\multicolumn{2}{|c|}{10 TTIs} 										\\
								& 	\multicolumn{2}{|c|}{(5 TTIs before and 5 TTIs after reassignment)} 				\\
		\hline
		Channel feedback delay 			& 	\multicolumn{2}{|c|}{1 ms} 										\\
		\hline
		\ac{enb} Antenna Configuration 		&	\multicolumn{2}{|c|}{4 \ac{tx} antennas}								\\
								&	\multicolumn{2}{|c|}{Cross-polarised 0.5$\lambda$ spacing, -45$^\circ/$,45$^\circ$ slants} 	\\
		\hline
		\ac{ue} Antenna Configuration 		&	\multicolumn{2}{|c|}{2 \ac{rx} antennas}								\\
								& 	\multicolumn{2}{|c|}{Cross-polarised 0.5$\lambda$ spacing, 0$^\circ/$,90$^\circ$ slants} 	\\
		\hline
		MIMO transmission scheme		&	\multicolumn{2}{|c|}{\ac{su-mimo}: single layer} 							\\
								&	\multicolumn{2}{|c|}{\ac{mu-mimo}: max. 2 \acp{ue}, 1 layer per \ac{ue}}			\\
		\hline
		Precoding Codebook			&	\multicolumn{2}{|c|}{Rel.8 4 \ac{tx} codebook}							\\
		\hline
		Initial Cell Selection 				&	\multicolumn{2}{|c|}{Maximum \ac{rsrp}}								\\
		\hline
		Feedback (To assigned cell)		&	\multicolumn{2}{|c|}{Subband \ac{cqi}, wideband \ac{pmi} for all \acp{ue}}			\\
		\hline
		Feedback (To centralised controller) 	&	\multicolumn{2}{|c|}{Wideband \ac{pmi} for each Target Neighbouring \ac{enb}}		\\
		\hline
		\ac{mu-mimo} $\Delta_{MUI}$		&	\multicolumn{2}{|c|}{0.05}										\\
		\hline
		\ac{ue} Scheduling 				&	\multicolumn{2}{|c|}{Proportional Fair \ac{sus}}							\\
		\hline
		\ac{sus} const $\epsilon$			&	\multicolumn{2}{|c|}{0.1}											\\
		\hline
		Traffic Model 				&	\multicolumn{2}{|c|}{Full Buffer}										\\
		\hline
		Inter-cell Interference model		&	\multicolumn{2}{|c|}{4 \ac{tx} \ac{su-mimo} with random \ac{pmi}}				\\
		\hline
		Feedback Overhead 			&	\multicolumn{2}{|c|}{31.15\%} 										\\
		\hline
	\end{tabular} }
	\vspace{-0.2cm}
\end{table}

%% file: SimulationResults.tex
\section{Results}

The following simulations were performed using the Matlab-based Vienna \ac{lte} System-Level simulator \cite{Ikuno2010}. 
To enable this investigation, we implemented as extensions \ac{mu-mimo} scheduling and transmissions with \ac{mmse}-\ac{irc} receivers, the investigated indoor and outdoor deployment scenarios, and small cell energy expenditure computation.

Monte Carlo simulations were performed as a series of snapshots which were averaged over a large number of iterations to generate the presented results. Once per snapshot the \emph{\ac{mu-mimo} across Small Cells} reassignment mechanism, described 
in section \ref{sec:mu-mimo_across_small_cells}, is carried out.
We assess what energy saving and spectral efficiency gains are achievable relative to the initial \ac{rsrp}-based assignments, which represent current practice in small cell deployments. 
We investigate how how these are affected by the number of \acp{ue} per cell, the reassignment parameter $\tau$, and the reassignment selection algorithm.
Vertical bars in all figures in this section correspond to the 95\% confidence interval.

\begin{figure}
	\centering
	\epsfig{file=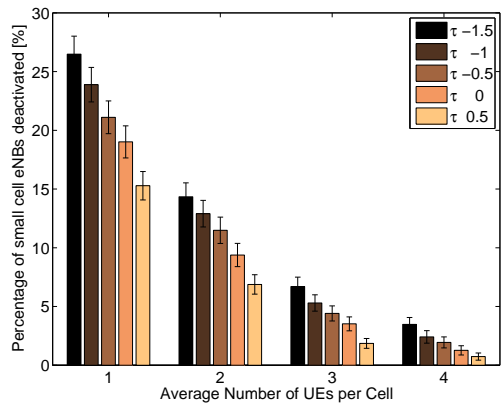,width=0.75\columnwidth}
	\caption{\footnotesize Percentage of \acp{enb} that can be switched to IDLE as a result of \ac{mu-mimo}-based cell reassignments for different reassignment thresholds $\tau$ in the indoor Dual Stripe scenario with \ac{dr} of 0.2.}
	\label{fig:deactivated_eNBs_dual_stripe}
	\vspace{-0.7cm}
\end{figure}

\begin{figure}
	\centering
	\epsfig{file=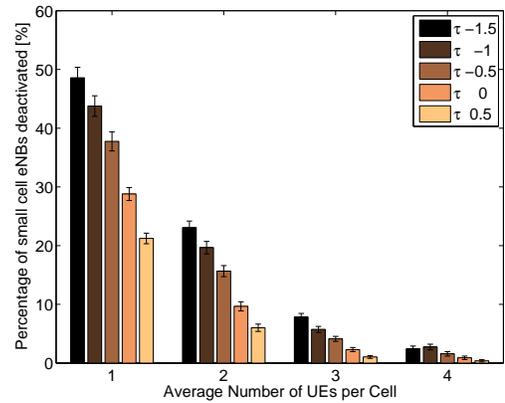,width=0.75\columnwidth}
	\caption{\footnotesize Percentage of \acp{enb} that can be switched to IDLE as a result of \ac{mu-mimo}-based cell reassignments for different reassignment thresholds $\tau$ in the outdoor Munich scenario.}
	\label{fig:deactivated_eNBs_Munich}
	\vspace{-0.6cm}
\end{figure}
	
\begin{figure}
	\centering
	\epsfig{file=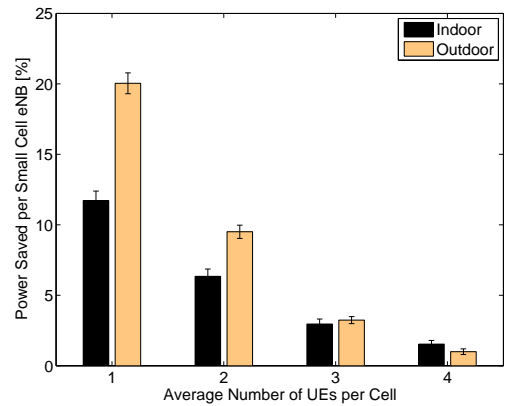,width=0.75\columnwidth}
	\caption{\footnotesize Average saved \ac{enb} power consumption through switching \acp{enb} to IDLE state as a result of \ac{mu-mimo}-based cell reassignments. $\tau = -1.5$.}
	\label{fig:power_saved_percent_DR}
	\vspace{-0.8cm}
\end{figure}

\vspace{-0.3cm}

\subsection{Energy Efficiency Gains}

Figures \ref{fig:deactivated_eNBs_dual_stripe} and \ref{fig:deactivated_eNBs_Munich} show the percentage of small cell \acp{enb} which are switched to a sleep mode as a result of the reassignment mechanism for the indoor Dual Stripe and outdoor Munich scenarios, respectively.
As can be seen in both figures the percentage of small cell \acp{enb} switched off decreases as the reassignment metric threshold $\tau$ increases. Higher $\tau$ corresponds to a stricter constraint on the foreseen reassignment spectral efficiency gains while increased leniency in $\tau$ results in a larger set of reassignable \acp{ue}, allowing more small cell \acp{enb} to be switched off.
While not shown, as $\tau$ continues to decrease, a point will be reached (roughly $\tau\!=\!-6$) where \ac{ue} reassignability becomes exclusively a function of \ac{ue} channel orthogonality and no longer of the neighbouring cell \acp{sinr}. In this case a \ac{ue} may be reassigned to any neighbouring \ac{enb} regardless of how far apart they may be, potentially resulting in dramatic decreases in spectral efficiency and inability to use \ac{mu-mimo} in the target neighbouring cell. Instead, in the figures we show a reasonable range of $\tau$ values for which this does not occur.

\vspace{-0.1cm}

The more \acp{ue} there are per \ac{enb} the less likely \textit{all} \acp{ue} in the cell will be reassignable. Further, if there are more \acp{ue} in the original cell it is more likely that suitable \ac{mu-mimo} pairs will already exist, reducing the number of considered \acp{ue}. This results in very low probabilities of a reassignment occurring for 4 \acp{ue} per cell, even for low $\tau$.

\vspace{-0.1cm}

Comparing Figures \ref{fig:deactivated_eNBs_dual_stripe} and \ref{fig:deactivated_eNBs_Munich}, the proportion of \acp{enb} deactivated is generally higher in the outdoor scenario. Figure \ref{fig:power_saved_percent_DR} presents the savings in small cell power consumption for the case of $\tau\!=\!-1.5$. We see that the power savings in the outdoor scenario are higher for few \acp{ue} per cell, while when there are many \acp{ue} per cell the power savings are comparable.

We see that in the outdoor case with 1 \ac{ue} per cell a 15\% reduction in the consumed power can be achieved. This corresponds to a 45\% energy saving on the 37\% percent the pico \acp{enb} which got deactivated or an average saving of 4.5W per pico cell. In the indoor case with the same parameters on average 2.5W per femto cell can be saved.

\subsection{Spectral Efficiency Gains}
\begin{figure}
	\centering
	\epsfig{file=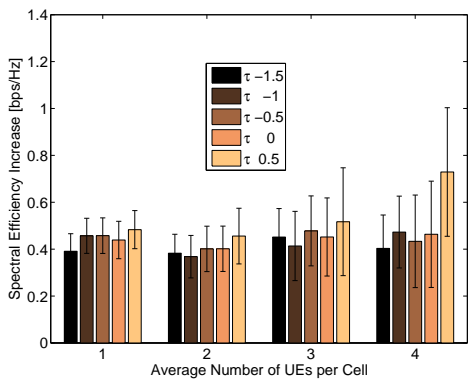,width=0.7\columnwidth}
	\caption{\footnotesize Increases in Spectral Efficiency resulting from reassignment, averaged across the Reassigned and Target \acp{ue} in the Dual Stripe indoor scenario.}
	\label{fig:spec_eff_increase_RUE_TUE_dual_stripe}
	\vspace{-0.5cm}
\end{figure}

\begin{figure}
	\centering
	\epsfig{file=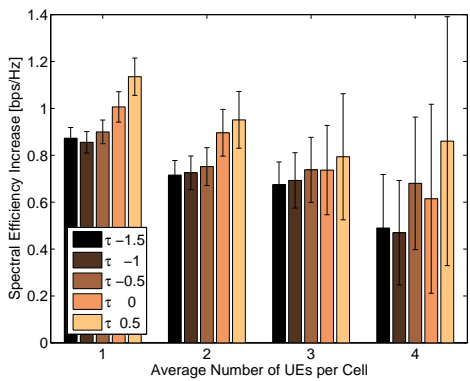,width=0.7\columnwidth}
	\caption{\footnotesize Increases in Spectral Efficiency resulting from reassignment, averaged across the Reassigned and Target \acp{ue} in the Munich outdoor scenario.}
	\label{fig:spec_eff_increase_RUE_TUE_Munich}
	\vspace{-0.7cm}
\end{figure}

Figures \ref{fig:spec_eff_increase_RUE_TUE_dual_stripe} and \ref{fig:spec_eff_increase_RUE_TUE_Munich} show the gains in spectral efficiency averaged over the \ac{ue} that gets reassigned and the Target \ac{ue} with which the Reassigned \ac{ue} is simultaneously served for \ac{mu-mimo} in the target neighbouring cell. Equivalently, these are the gains in spectral efficiency of the enabled \ac{mu-mimo} usage over the pre-reassignment \ac{su-mimo2} usage, and are a function of the original and target \ac{enb} \acp{sinr}, before and after reassignment.

As expected, higher reassignment threshold $\tau$ generally corresponds to higher spectral efficiency increase; however, the effect of $\tau$ is not strong.
Due to it not being possible to know prior to reassignment which neighbouring cells will also be deactivated, $\tau$ cannot take into account gains in \ac{sinr} from a neighbouring \ac{enb} switching to sleep state (thus removing the inter-cell interference it causes). Further, lower values of $\tau$ mean that more reassignments can be performed, allowing more small cells to be deactivated, and resulting in more reductions in interference. 
This also explains why the increases in spectral efficiency tend to exceed the values of $\tau$.

We see that the gains in spectral efficiency in the outdoor scenario are higher than in the indoor scenario. This is due to the lack of walls between neighbouring small cell \acp{enb} on the same street. In the indoor scenario there are walls between all \acp{enb}, meaning that the minimum difference in \ac{sinr} between two neighbouring cells is higher. This further explains why the proportion of \acp{enb} that could be deactivated was higher in the outdoor scenario in Figure \ref{fig:deactivated_eNBs_Munich}.

%% file: ConclusionsAndFutureWork.tex
\vspace{-0.3cm}

\section{Conclusions}

In this work we have demonstrated the combined use of \ac{mu-mimo}-based \ac{ue} reassignments and centralised control of small cell sleep states to achieve simultaneous increases in spectral efficiency and reductions in energy consumption.
We compare the achievable gains in two small cell scenarios, one indoor residential apartment block scenario (Dual Stripe) and one commercial city centre outdoor scenario (Stachus Square, Munich). We found that it was possible to switch over 25\% (indoor) and 35\% (outdoor) of small cells to a sleep state whilst still achieving considerable gains in spectral efficiency. Based on the energy consumption model used \cite{Imran2010} these correspond to power savings of 12\% and 15\% respectively.
\vspace{-0.2cm}